\documentclass{osa-article}

\journal{oe}

\usepackage{verbatim}
\usepackage{graphicx}
\usepackage{amsmath}
\usepackage{afterpage}
\usepackage{layouts}
\usepackage[normalem]{ulem}

\DeclareMathOperator{\e}{e}
\usepackage{hyperref}
\hypersetup{pdfstartview={FitH},pdfpagemode={UseNone},
	colorlinks,linkcolor=blue, citecolor=blue, urlcolor=blue,
	bookmarksopen=true, pdfnewwindow=true}

\usepackage[all]{hypcap}

\begin{document}
\title{Nonlinear transition between PT-symmetric and PT-broken modes in coupled fiber lasers}

\author{Sergey~V.~Suchkov\authormark{1,2,*},
Dmitry~Churkin\authormark{1}, and
Andrey~A.~Sukhorukov\authormark{2}}

\address{
\authormark{1}Novosibirsk State University, 2 Pirogova St., Novosibirsk, 630090, Russia\\
\authormark{2}Nonlinear Physics Centre, Research School of Physics, Australian National University, Canberra, ACT 2601, Australia}

\email{\authormark{*}sergey.v.suchkov@gmail.com} 



\begin{abstract}
We present a systematic analysis of the stationary regimes of nonlinear parity-time (PT) symmetric laser composed of two coupled fiber cavities. We find that power-dependent nonlinear phase shifters broaden regions of existence of both PT-symmetric and PT-broken modes, and can facilitate transitions between modes of different types. We show the existence of non-stationary regimes and demonstrate an ambiguity of the transition process for some of the unstable states. We also identify the presence of higher-order stationary modes, which return to the initial state periodically after a certain number of round-trips.
\end{abstract}

\section{Introduction}

The concept of parity-time (PT) symmetry is extensively used in the design of diverse optical devices with balanced gain and loss, which provide new possibilities for effective signal manipulation.
Specially tailored symmetric distributions of the gain and loss regions, as well as the refractive index profile, can support optical eigenmodes which energy is conserved due to a balance of gain and loss, while the modes exhibit amplification or attenuation after crossing a bifurcation point, the so-called PT-symmetry breaking threshold.
The total average power can be conserved in the PT-symmetric phase, however there might appear transitional power variations and oscillations related to the non-orthogonality of the interfering eigenmodes.
Thus PT-symmetric systems combine properties of conservative and active systems and can be used for light amplification, filtering and switching~\cite{Christodoulides:2018:ParityTime, El-Ganainy:2018-11:NPHYS, Konotop:2016-35002:RMP, Suchkov:2016-177:LPR}. PT-symmetric systems posses an exceptional point, which separates PT-symmetric and broken phases and plays an important role in linear and nonlinear dynamics~\cite{Ozdemir:2019-783:NM, El-Ganainy:2018-11:NPHYS}. The PT-symmetric structures belong to a broader class of pseudo-Hermitian systems with an entirely real spectrum, which was proposed and extensively studied in recent years~\cite{Mostafazadeh:2002-205:JMP, Mostafazadeh:2010-1191:IJGM, Suchkov:2016-65005:NJP}.

The idea to use PT-symmetry in laser systems is attracting increasing attention. A couple of active and passive fiber rings was studied in~\cite{Regensburger:2012-167:NAT}, where the same dynamics as in a PT-symmetric synthetic lattice were demonstrated. The PT-broken regime can be employed to realize single-mode lasing in multi-mode systems. The PT lasers were experimentally demonstrated in microrings~\cite{Feng:2014-972:SCI, Hodaei:2014-975:SCI, Liu:2017-15389:NCOM, Ren:2018-27153:OE} and fiber resonators~\cite{Jahromi:2017-1359:NCOM}. A single transverse mode operation in coupled microring lasers was demonstrated near the exceptional point~\cite{Hodaei:2016-494:LPR}, and enhanced sensitivity~\cite{Ren:2017-1556:OL, Hodaei:2017-187:NAT, Smith:2019-34169:OE} was realized.
In the coupled microdisk quantum cascade lasers, the reversal of generated power dependence was identified in the vicinity of exceptional points~\cite{Brandstetter:2014-4034:NCOM}, where spontaneous emission was enhanced~\cite{Pick:2017-12325:OE}, and these points were observed directly in photonic-crystal lasers~\cite{Kim:2016-13893:NCOM}.
Additionally, a realization of a PT symmetry-based mode-locking~\cite{Longhi:2016-4518:OL} was theoretically proposed and it was shown that a non-Hermitian phase transition can be observed in the frequency domain~\cite{Longhi:2019-1190:OL}. Furthermore, lasing and anti-lasing can be realized in a single cavity~\cite{Wong:2016-796:NPHOT}.

In fiber systems, it was shown that, despite strong phase stochasticity, pure PT-symmetry phase can be still observed in long optical fiber network incorporating semiconductor amplifiers~\cite{Jahromi:2017-1359:NCOM}. It was predicted that PT-symmetric coupled fiber-loop lasers can exhibit bi-stable dynamics combined with a lower lasing power threshold~\cite{Smirnov:2018-A18:PRJ}. In this work, we study the effect of nonlinear phase modulation on the dynamics of the PT-symmetric coupled fiber-loop laser and identify the transition dynamics between the PT-symmetric and broken phases.




\section{Nonlinear PT-symmetric fiber loop laser}

\begin{figure}[htb]
  \centerline{\includegraphics[width=0.9\columnwidth]{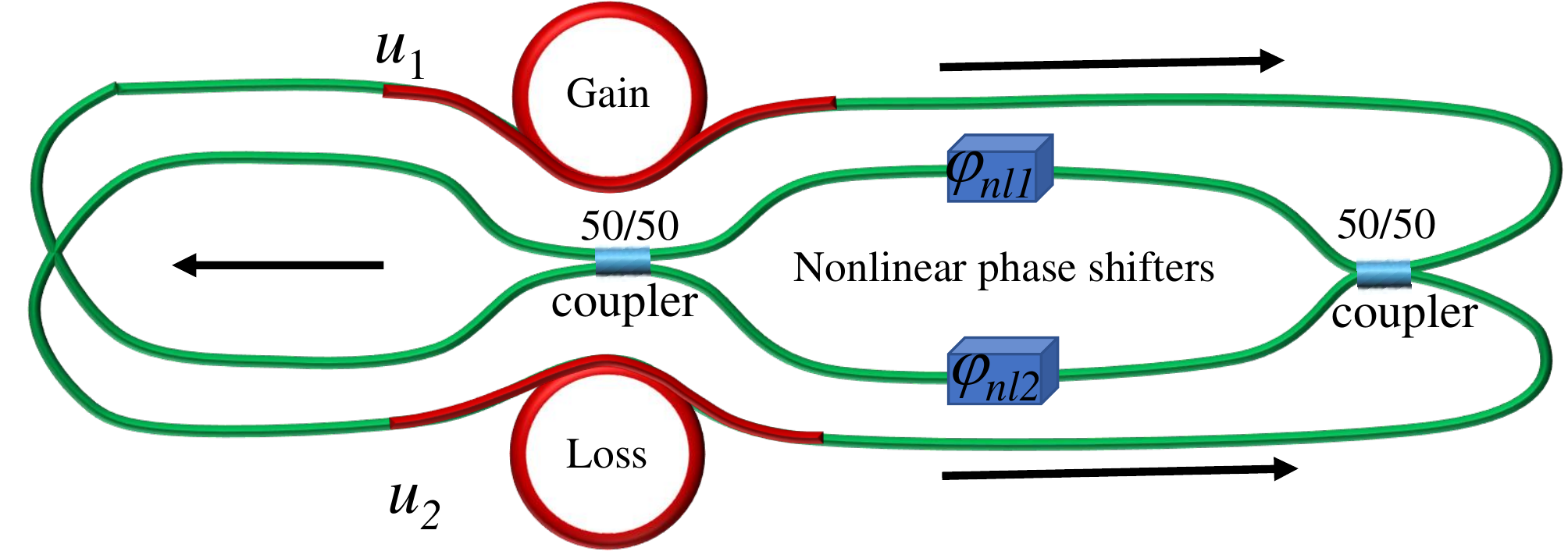}}
  \caption{\label{FigScheme}
Schematic diagram of the PT-symmetric laser composed of two coupled fiber ring cavities with gain and loss. Red sectors of the fiber indicate gain and loss regions, arrows show the direction of propagation.}
\end{figure}

We consider a system comprised of two identical fiber cavities, one of them is pumped (i.e. active), another one is passive, as schematically illustrated in Fig.~\ref{FigScheme}. The cavities are coupled by means of phase shifters providing a phase shift of a propagating signal. In addition, the cavities are cross-coupled, which  facilitates the PT-symmetry of the system. In this work, we systematically investigate the effect of power-dependent nonlinear phase shift and identify distinct phenomena compared to the previously analyzed linear phase shifters~\cite{Smirnov:2018-A18:PRJ}.

A signal round trip can be described by matrix operator ${\bf L}$, where each sub-matrix corresponds to a couple of controlling elements placed in both loops of the laser setup a shown in Fig.\ref{FigScheme}: gain and loss, nonlinear phase shifters, 50-50 couplers, and cross-coupling.
\begin{equation}\label{eq:Transfer}
  {\bf L}=
  \left(
  \begin{array}{cc}
    0 & 1 \\
    1 & 0
  \end{array}
  \right)\left(
  \begin{array}{cc}
    \frac{1}{\sqrt{2}} & \frac{i}{\sqrt{2}} \\
    \frac{i}{\sqrt{2}} & \frac{1}{\sqrt{2}}
  \end{array}
  \right)\left(
  \begin{array}{cc}
    \e^{i\varphi_{{\rm nl}1}} & 0 \\
    0 & \e^{i\varphi_{{\rm nl}2}}
  \end{array}
  \right)\left(
  \begin{array}{cc}
    \frac{1}{\sqrt{2}} & \frac{i}{\sqrt{2}} \\
    \frac{i}{\sqrt{2}} & \frac{1}{\sqrt{2}}
  \end{array}
  \right)\left(
  \begin{array}{cc}
    \e^{g_1} & 0 \\
    0 & \e^{g_2 }
  \end{array}
  \right),
\end{equation}
where $g_2 < 0$ and $g_1 > 0$ are the loss and gain coefficients, respectively, $\varphi_{{\rm nl}1}$ and $\varphi_{{\rm nl}2}$ are the nonlinear phase shifts provided by phase shifters in the active and passive loops. PT-symmetry of the considered system follows from the fact that the operator
\begin{equation}\label{eq:PTtr}
  {\bf L}_{\rm PT}=\left(
  \begin{array}{cc}
    \e^{g_1/2} & 0 \\
    0 & \e^{g_2/2 }g
  \end{array}
  \right){\bf L}\left(
  \begin{array}{cc}
    \e^{-g_1/2} & 0 \\
    0 & \e^{-g_2/2 }
  \end{array}
  \right)
\end{equation}
is PT invariant for $g_1+g_2=0$, as was shown in~\cite{Smirnov:2018-A18:PRJ}, however, the form Eq.~\eqref{eq:Transfer} is more convenient for analysis.

In general, nonlinear dynamics is a comprehensive process, where stationary states play an important role and may serve as attractors.
Indeed, in the previously considered case of linear phase shifters and nonlinear gain saturation~\cite{Smirnov:2018-A18:PRJ}, it was found that any initial input leads to one of the stationary states.
A stationary mode, $\bf{u}$, is determined as $|{\bf L}({\bf u}){\bf u}|=|{\bf u}|$, which means that $\bf{u}$ is an eigenvector of $\bf{L(u)}$ with an eigenvalue $|\mu|=1$. Following the method proposed in~\cite{Smirnov:2018-A18:PRJ} and taking into account Eq.~\eqref{eq:PTtr}, we identify the eigenvalues and eigenvectors of the operator ${\bf L}$ as
%
\begin{equation}\label{eq:mu}
\begin{split}
  \mu_{\pm}  = i \e^{ \tilde{g} + i \tilde{\varphi}_{\rm nl} }\left(
                      {\rm cosh}( \Delta g / 2 ) \cos( \Delta\varphi_{\rm nl} / 2 ) \pm \sqrt{ {\rm cosh}^2( \Delta g / 2 ) \cos^2( \Delta\varphi_{\rm nl} / 2 ) - 1 }\right),
\end{split}
\end{equation}
\begin{equation}\label{eq:modesPT}
   \begin{split}
   {\bf u}&=
   \left(\begin{array}{c}
    u_1 \\
    u_2
  \end{array}\right) =A
  \left(\begin{array}{c}
    1 \\
    \exp(- i \nu_\pm +\Delta g/2)
  \end{array}\right), \\
  \mbox{ where } \nu_\pm &= - i \log\left[\frac{ -i\e^{-\tilde g-i\tilde{\varphi}_{\rm nl}}\mu_\pm - e^{-\Delta g / 2} \cos\left(\Delta\varphi_{\rm nl}/2 \right)}{
                            \sin\left(\Delta\varphi_{\rm nl}/2 \right) }\right],
                            \end{split}
\end{equation}
and $A$ is an amplitude. Here we denote $\tilde{g} \equiv (g_1+g_2) / 2$, $\Delta g \equiv g_1-g_2$, $\tilde{\varphi}_{\rm nl} \equiv (\varphi_{\rm nl1}+\varphi_{\rm nl2}) / 2$ is a common phase shift, and $\Delta\varphi_{\rm nl} \equiv \varphi_{\rm nl1} - \varphi_{\rm nl2}$ is a relative phase shift. Since we consider power-dependent phase shifters, $\Delta\varphi_{\rm nl}$ depends on eigenvector ${\bf u}$. Moreover, in a real laser system gain usually exhibits saturation with power growth~\cite{Ge:2016-24889:SRP}. Thus we approximate it by the following relation $g_1=g_0/(1+\alpha P_1)-g_h$, where $P_1$ is the signal intensity in the first loop at the current round trip, $g_h$ is the value of loss in the gain element when intensity is high, $g_0$ determines the gain at low powers and $\alpha$ defines an inverse of the characteristic gain-saturation power.

\begin{figure}[htb]
  \centerline{\includegraphics[width=0.8\columnwidth]{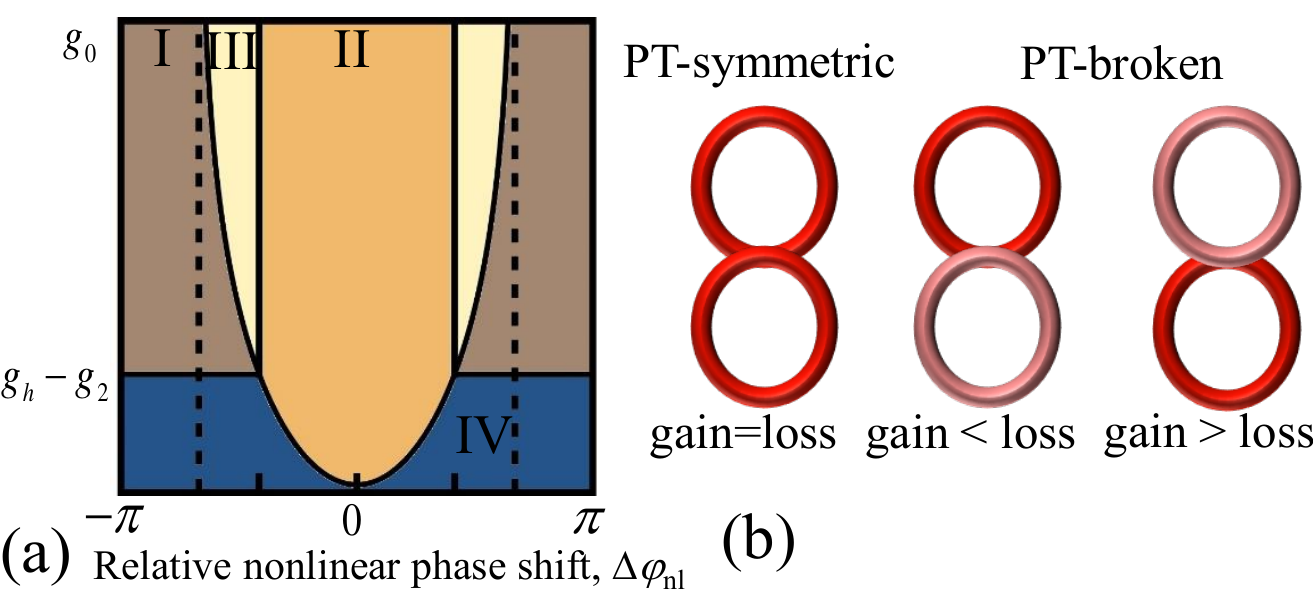}}
  \caption{\label{Fig1}
 (a) Parameter regions of existence of stationary modes. Brown (I) -- a couple of PT-symmetric modes, Orange (II) -- only one PT-broken mode with $|\mu_+|=1$  ($|\mu_-|<1$), Yellow (III) -- a couple of PT-symmetric and one PT-broken modes with $|\mu_-|=1$ co-exist ($|\mu_+|>1$). (b) Schematically shown averaged power distribution in stationary regimes for the PT-symmetric case - power distributed equally between loops, and PT-broken cases - power mostly concentrated in active (orange region)  or passive (yellow region) loops, respectively.
}
\end{figure}

We now analyze the balance condition for a stationary mode, $|\mu_+|=1$ and/or $|\mu_-|=1$,
\begin{eqnarray}
\label{g1r}
\e^{\tilde{g}}\left|\cosh(\Delta g/2)\cos(\Delta\varphi_{\rm nl}/2)\pm\sqrt{\cosh(\Delta g/2)^2\cos(\Delta\varphi_{\rm nl}/2)^2-1} \right|=1,
\end{eqnarray}
where according to the gain saturation
\begin{eqnarray}
\label{gs}
 g_1 = g_0 (1+\alpha A^2)^{-1} - g_h.
\end{eqnarray}
These relations place a constraint on the gain parameter $g_1$, relative phase shift $\Delta\varphi_{\rm nl}$, and the amplitude $A$.
Using these restrictions we calculate the parameter regions where different types of stationary modes of ${\bf L}$ exist and plot them in Fig.~\ref{Fig1}~(a). For definiteness we set $g_2=-0.7$ and $g_h=0.05$ as in Ref.~\cite{Smirnov:2018-A18:PRJ}.
If the gain at low powers, $g_0$, is less than critical value, $g_h$, lasing does not occur in the system (blue region IV in Fig.~\ref{Fig1}~(a)). In this case Eqs.~\eqref{g1r}-\eqref{gs} can not be satisfied for any $A\geq 0$.
When $g_0\geq g_h-g_2$,  there are a couple of PT-symmetric modes with $|\mu_{+}|=|\mu_{-}|=1$ and $g_1+g_2=0$ if $|\Delta\varphi_{\rm nl}|>2\arccos(\rm{sech}[g_2])$. The latter expression is a PT-symmetry breaking threshold as defined in~\cite{Smirnov:2018-A18:PRJ}. This case spans brown (I) and yellow (III) regions in Fig.~\ref{Fig1}~(a). For the parameter range $|\Delta\varphi_{\rm nl}|<2\arccos(\rm{sech}(g_2))$ there is only one PT-broken mode with $|\mu_+|=1$ for some $g_1<-g_2$, while the other mode exponentially decays during propagation since $|\mu_-|<1$ (orange region II).

We note that for $\arccos(\rm{sech}(g_2))<|\Delta\varphi_{\rm nl}|<2\arccos(\e^{g_2})$ there is a stationary mode with $|\mu_-|=1$ for a specific range of $g_1>-g_2$, while another mode has $|\mu_+|>1$ (yellow region II) and it grows exponentially. In this region, both types of modes co-exist, although for different values of $g_1$ and $A$. In Fig.~\ref{Fig1}~(b) we schematically present averaged power distribution in the cavities corresponding to different types of stationary modes: PT-symmetric case - equal power distribution, PT-broken case - power preferably concentrated in the first or second loop. 
\begin{figure}[htb]
  \centerline{\includegraphics[width=0.8\columnwidth]{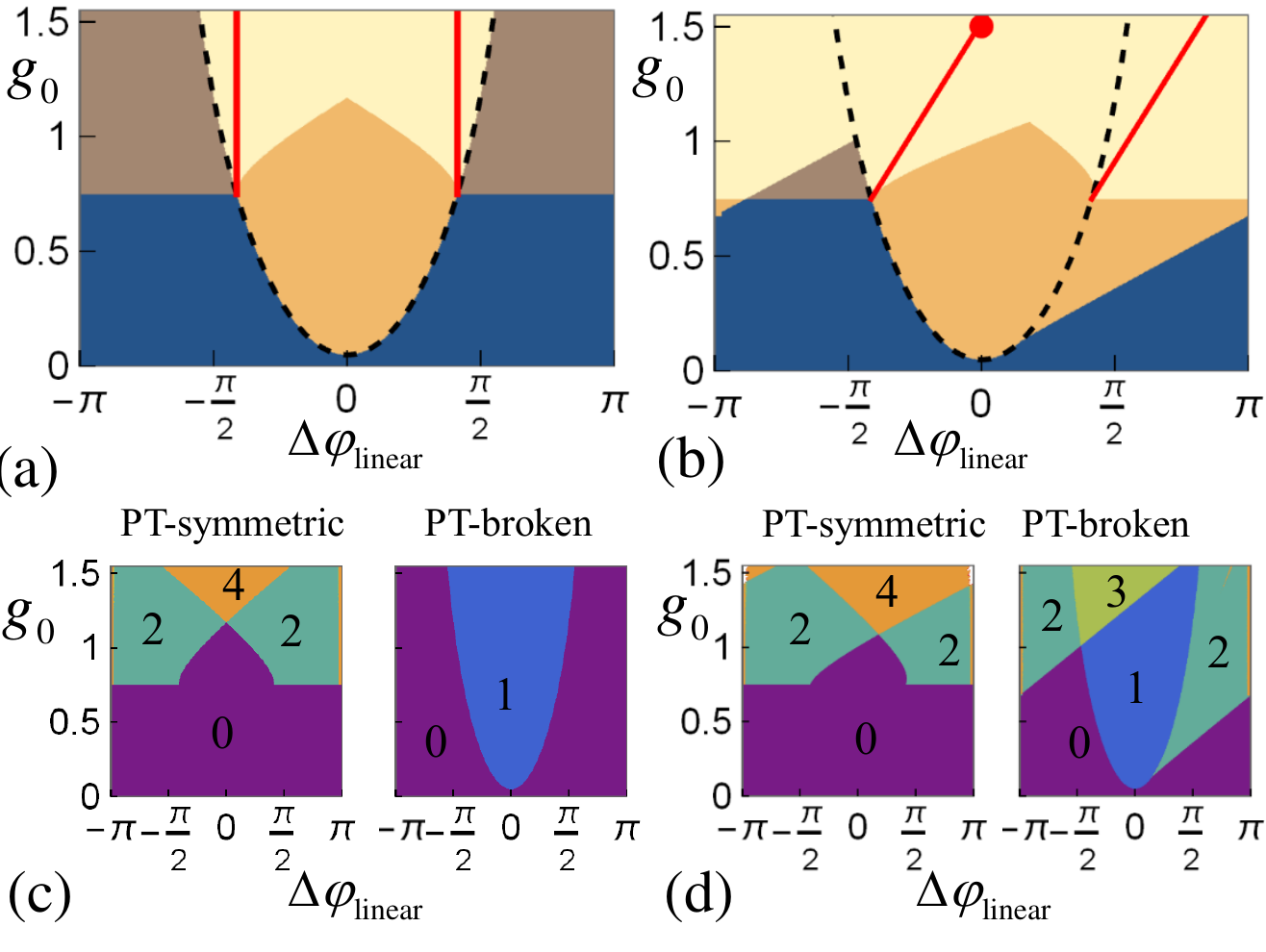}}
  \caption{\label{Fig2}
  Parameter regions where stationary modes exist. Blue - no modes exist, Grey - only PT-symmetric modes exist, Orange - only PT-broken modes exist, Light Brown - modes of both types co-exist. Red lines show PT-symmetry breaking threshold. (a, c) $\gamma_1=\gamma_2=2$; (b, d)  $\gamma_1=2$, $\gamma_2$=3. (c,d) Number of PT-symmetric and PT-broken modes corresponding panels (a,b), respectively.}
\end{figure}
As discussed above, the relative phase shift $\Delta\varphi_{\rm nl}$ and the gain at low powers $g_0$ determine the type of a stationary regime in which the system operates in a particular case. However, $\Delta\varphi_{\rm nl}$ is power dependent and it consists of the linear (constant) and nonlinear parts as
\begin{equation}
\label{nlp}
	\Delta\varphi_{\rm nl}=\Delta\varphi_{\rm linear}+\gamma_1 \hat{P}_1-\gamma_2 \hat{P}_2, 	
\end{equation}
where $\gamma_{1,2}$ are the nonlinear coefficients and $\hat{P}_{1,2}$ are signal powers inside the phase shifters in the active and passive loops, respectively, $\Delta\varphi_{\rm linear}$ is a constant relative phase shift produced by the phase shifters when the signal power is small. This means that a particular laser setup can demonstrate a variety of stationary regimes corresponding to different signal powers.
The nonlinearity can effectively "mix" the different regions shown in Fig.~\ref{Fig1}, enabling more than three stationary regimes to co-exist in specific fixed laser configurations. Taking into account Eqs.~\eqref{eq:modesPT}-\eqref{nlp} we find that
\begin{multline}
	\label{eq:phase}
	\Delta\varphi_{\rm nl}=\Delta\varphi_{\rm linear}+\frac{g_0-g_1(\Delta\varphi_{\rm nl})-g_h}{g_1(\Delta\varphi_{\rm nl})+g_h}\\
	\times\left(|A_1(\Delta\varphi_{\rm nl})|^2 \frac{\gamma_1-\gamma_2}{\alpha} +\gamma_2/\alpha(|A_1(\Delta\varphi_{\rm nl})|^2-|A_2(\Delta\varphi_{\rm nl})|^2)\right),
\end{multline}
where $A_1$ and $A_2$ are determined through eigenvectors defined in Eq.~\eqref{eq:modesPT}:
\begin{eqnarray}
	\left(\begin{array}{c}
	A_1(\Delta\varphi_{\rm nl})\\
    A_2(\Delta\varphi_{\rm nl})
	\end{array}\right)=\frac1A\left(
  \begin{array}{cc}
    {1}/{\sqrt{2}} & {i}/{\sqrt{2}} \\
    {i}/{\sqrt{2}} & {1}/{\sqrt{2}}
  \end{array}
  \right)\left(
  \begin{array}{cc}
    e^{g_1(\Delta\varphi_{\rm nl})} & 0 \\
    0 & e^{g_2 }
  \end{array}
  \right)\left(
  \begin{array}{c}
  u_1\\
  u_2
  \end{array}
  \right).
\end{eqnarray}

We numerically solve Eq.~\eqref{eq:phase} to determine $\Delta\varphi_{\rm nl}$ for the chosen $\Delta\varphi_{\rm linear}$ and $g_0$ and thereby identify the mode type according to the region $\Delta\varphi_{\rm nl}$ belongs to. The results are presented in Fig.~\ref{Fig2} for $g_2=-0.7$, $g_h=0.05$, and $\alpha=2$. In panel (a) we show the regions where a stationary mode exists for $\gamma_1=\gamma_2=2$, while in (b) $\gamma_1=2$ and $\gamma_2=3$. Blue color shows the region (IV) where no lasing modes exist, Grey (I) -- only PT-symmetric modes exist, Orange (II) -- only PT-broken modes exist, Light Brown (III) -- modes of both types co-exist. Red lines indicate the PT-symmetry breaking threshold where both eigenvalues and eigenvectors collapse. In panels~(c) and~(d) we indicate the total number of PT-symmetric and PT-broken stationary modes corresponding to the parameters in~(a) and~(b), respectively.

We observe in Figs.~\ref{Fig2}~(a, b) that the nonlinear phase modulation broadens the regions of existence of both the PT-symmetric and PT-broken modes, and accordingly increases the overlap region where the modes of different symmetries can coexist. We also note that the total number of stationary modes of both types increases with the gain at low powers growth $g_0$ as well.

Another interesting aspect is that in the case of identical nonlinear coefficients in the two loops, $\gamma_1=\gamma_2$, the region of stationary PT-broken modes coincides with one for the case of linear phase shifters considered in~\cite{Smirnov:2018-A18:PRJ}.
Indeed, for PT-broken modes $|A_1|=|A_2|$ and accordingly from Eq.~\eqref{eq:phase} we obtain $\Delta\varphi_{\rm nl}=\Delta\varphi_{\rm linear}$.
Thus, nonlinearity does not affect the stationary PT-broken modes and consequently, there cannot be more than one stationary PT-broken mode for any fixed laser parameters.


\section{Transitions in PT-symmetric fiber laser}
\begin{figure}[htb]
  \centerline{\includegraphics[width=\columnwidth]{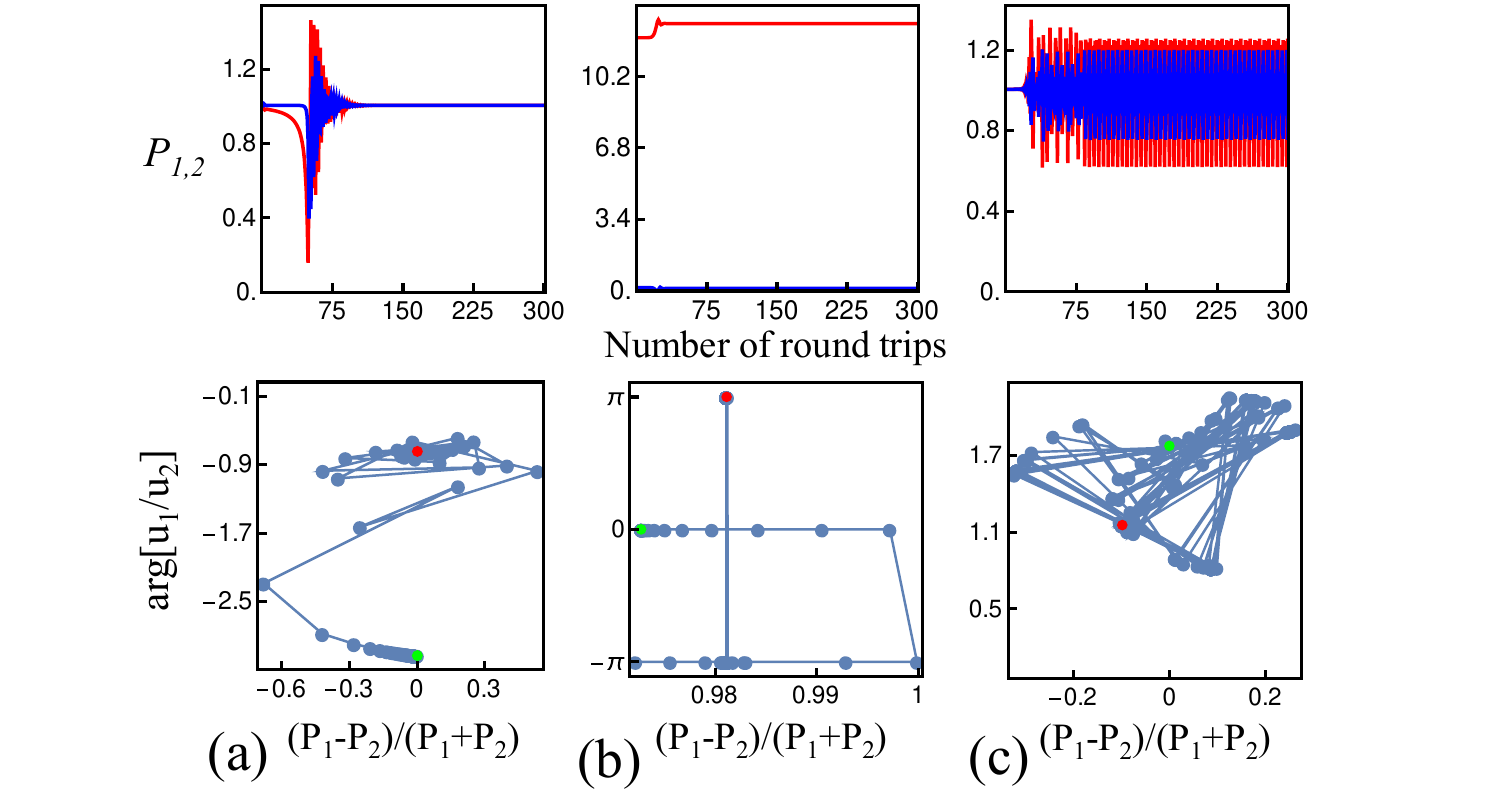}}
  \caption{\label{Fig3}
    Nonlinear transitions in the PT-symmetric fibre laser. Top row - laser dynamics in active (red) and passive (blue) loops, bottom row - corresponding trajectories in the phase plane. Green and red dots show where the trajectories start and finish, respectively. Parameters are $\Delta\varphi_{\rm linear}=0$, $g_0=1.5$ as indicated by the red point in Fig.~\ref{Fig2}~(b). (a) ${\bf u}_0=(0.71, -1.42 + 0.014 i)$, (b) ${\bf u}_0=(3.45, 0.58)$, (c) ${\bf u}_0=(0.70, -0.29 - 1.39 i)$.
  }
\end{figure}

We now investigate the dynamical transitions between a variety of co-existing stationary modes that we have identified above. To demonstrate these processes we apply initial conditions corresponding to different stationary modes with small perturbations. In simulations, we track the powers
in the middle of the gain/loss elements, i.e. $(P_1^{(n)},P_2^{(n)})=(\e^{g_1} |u_{1}^{(n)}|^2,\e^{g_2} |u_{2}^{(n)}|^2)$, where $n$ is the round-trip number. Then, PT-symmetric modes correspond to equal power distribution between the loops ($P_1=P_2$), while in PT-broken regime the powers are different.
In Fig.~\ref{Fig3}, top row, we show characteristic examples of the transition dynamics for the parameters marked by the red point in Fig.~\ref{Fig2}~(b), where four PT-symmetric and three PT-broken stationary modes can exist. The bottom row in Fig.~\ref{Fig3} shows the evolution in a phase space. A stable stationary state would be just a point in the phase space, while transition process is a trajectory from one point to another.
In Fig.~\ref{Fig3}~(a) an unstable PT-symmetric mode transits to a stable PT-symmetric one, in (b) a transition regime between two PT-broken modes is presented. Another interesting observation is the existence of non-stationary regimes, see an example in Fig.~\ref{Fig3}~(c).
We find  that this regime corresponds to a higher-order stationary mode, where the mode profile is restored after $N$ round-trips,
i.e. ${\bf L}^N{\bf u}=\exp(i \beta) {\bf u}$, where $\beta$ is a real-valued phase coefficient. For the case in Fig.~\ref{Fig3}~(c), the periodicity is $N=4$.

To better understand the transition processes we use the property that the relative phase shift between the mode amplitudes in active and passive loops conserves its sign after each round trip, see~\cite{Smirnov:2018-A18:PRJ} for a general proof applicable to both linear and nonlinear phase shifters. Specifically,
${\rm sign} \sin(J_n)={\rm sign}\sin(J_0)$ for any initial condition ${\bf u^{(0)}}$, where we denote $J_n=(\arg(u_1^{(n)})-\arg(u_2^{(n)}))$. Then, from Eqs.~\eqref{eq:mu}-\eqref{eq:modesPT} it follows that a PT-symmetric mode with ${\rm sign}\sin(J)=1$ cannot transit to a different PT-symmetric mode with ${\rm sign}\sin(J)=-1$. However, for PT broken modes $J=0,\pm \pi$ and there could be transitions of different types under small perturbations. In the linear regime, any PT-symmetric mode with ${\rm sign}\sin(J)=1$ has it counterpart with ${\rm sign}\sin(J)=-1$ and vice versa, however, there are always only two PT-symmetric modes and therefore a transition between them is prohibited~\cite{Smirnov:2018-A18:PRJ}. In contrast, the nonlinearity of phase shifters can facilitate a transition between all types of modes. Furthermore, the nonlinear fiber laser possesses multi-stability. Doe example, for the laser parameters considered in Fig.~\ref{Fig3} we find three stable fundamental stationary modes, as well as two stationary modes of higher order $N>1$.

\begin{figure}[htb]\label{FigLast}
  \centerline{\includegraphics[width=\columnwidth]{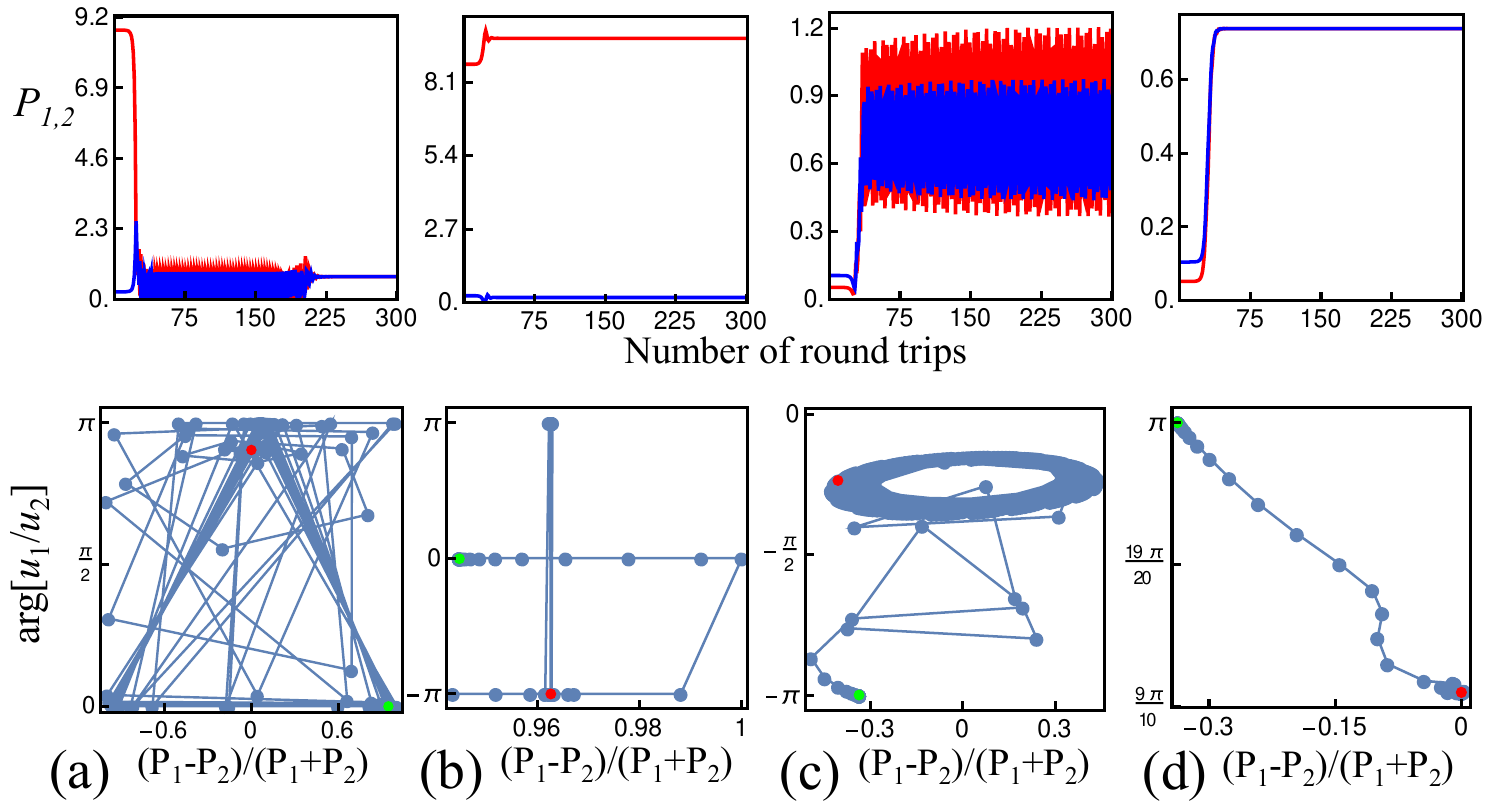}}
  \caption{\label{Fig4}
  Unstable transitions in the PT-symmetric fiber laser. Top row~-- laser dynamics in active (red) and passive (blue) loops, bottom row~-- corresponding trajectories in the phase plane. Green and red dots show where the trajectories start and finish, respectively.  The input signal is $(u_1 \exp[i\phi],u_2)$, where $(u_1,u_2)$ is a stationary mode and $\phi$ is a small perturbation. (a) and (b) $(u_1,u_2)=(2.92971, 0.708525)$, $\phi=\pm10^{-6}$, (c) and (d) $(u_1,u_2)=(0.124196, -0.458554)$, $\phi=\pm10^{-6}$, respectively. Other parameters are $g_0=1.3$, $\Delta\varphi_{\rm linear}=-1.5$, $\gamma_1=2$, $\gamma_2=3$.
  }
\end{figure}

We also noticed that some unstable stationary states demonstrate different dynamics depending on the perturbation, i.e. they can transit to at least two different regimes. In Fig.~\ref{FigLast} we show the dynamics of stationary modes where we introduce a perturbation $\phi$ as $(u_1 \exp[i\phi],u_2)$. This small perturbation shifts the mode upper or lower in the phase plane (as shown in the bottom row of Fig.~\ref{FigLast}), which for a PT-broken mode located on the "border" between upper and lower phase semi-plane, leads to a transition to different regimes with ${\rm sign}\sin(J_n)={\rm sign}\sin(\phi)$. Parameters are indicated in the figure caption.
Interestingly, we observe in Fig.~\ref{FigLast}~(a) that there can be several transitions during the evolution, e.g. from a PT-broken mode to a higher-order mode and then to a PT-symmetric mode. On the other hand, Fig.~\ref{FigLast}~(c) illustrates a transition to a stable higher-order mode, which is a distinct feature due to the nonlinearity of phase shifters.

\section{Conclusion}

In this paper, we revealed distinct features of the modes and their dynamical transitions in a nonlinear PT-symmetric fiber-loop laser. We developed a semi-analytical approach to systematically find all fundamental stationary modes. We demonstrated that the power-dependent phase shifts can mix the regions of existence of PT-symmetric and PT-broken stationary modes. We found that the higher is the gain at low powers ($g_0$), the larger number of stationary modes can co-exist simultaneously. We identified dynamical transitions between modes of the same or different PT symmetries, which can exhibit multi-stability. We also observed in numerical modeling non-stationary regimes that could not occur in the case of linear phase shifters, including higher-order stationary modes with different periodicity.
We anticipate that our findings will be useful for practical realization of lasing mode engineering and switching in PT-symmetric fiber-loop lasers, since Kerr-type nonlinearity naturally appears in fibers.

\section*{Funding}
A.A.S. and S.V.S. acknowledge support by the Australian Research Council (ARC) (DP160100619 and DP190100277). S.V.S. also thanks Foundation for the Advancement of Theoretical Physics and Mathematics (BASIS) (18-1-3-39-1). D.V.C. acknowledges support by Russian Foundation for Basic Research (19-0200633).
\bibliography{db_PTlaser}





\end{document}